\documentclass[aps,prd,preprint,groupedaddress,showpacs]{revtex4}

\usepackage[ansinew]{inputenc}
\usepackage{amsfonts}
\usepackage{color}

\usepackage{setspace}
\usepackage{amsmath}

\newcommand{\be}{\begin{equation}}
\newcommand{\ee}{\end{equation}}
\newcommand{\ba}{\begin{eqnarray}}
\newcommand{\ea}{\end{eqnarray}}

\newcommand{\mb}{\text}

\newcommand{\fr}{\frac}
\newcommand{\dl}{\partial}

\newcommand{\re}{\textrm{e}}

\begin{document}

\title{Application of Quantum Darwinism\\ to Diffusion during Cosmic Inflation}

\author{Nicol\'{a}s F. Lori}

\affiliation{IBILI, Universidade de Coimbra, 3000-354 Coimbra, Portugal}

\email{nflori@fmed.uc.pt}

\author{Alex H. Blin}

\affiliation{Departamento de F\'{\i}sica, Universidade de Coimbra, 3004-516
Coimbra, Portugal}

\email{alex@fis.uc.pt}


\begin{abstract}

A baby-universe model of cosmic inflation is analyzed
using quantum Darwinism. In this model cosmic inflation can be
approximated as Brownian motion of a quantum field, and quantum
Darwinism implies that decoherence is the result of  quantum Brownian
motion of the wave function. The quantum Darwinism approach to
decoherence in the baby-universe cosmic-inflation model yields
the decoherence times of the baby-universes. The result is the
equation relating the baby-universe's decoherence time with the
Hubble parameter. A brief discussion
of the relation between Darwinism and determinism is provided
in the Appendix.

\end{abstract}


\maketitle

\section{Introduction}

In this work a possible relation between universe creation during
cosmic inflation and quantum Darwinism is proposed. Linde's approach
to the Big Bang \cite{1} indicates that the creation of a universe
from nothing occurs in a Brownian motion-like process. Zurek's
quantum Darwinism approach to quantum mechanics \cite{2} indicates
that the reason for quantum indeterminacy is that the existence
of a state is directly related to that state's capacity to transmit
information about itself, and that this capacity is related to
diffusion for the case of quantum Brownian motion \cite{2}.

The approaches of Linde and Zurek are examples of two diverging
approaches to quantum physics. There is no disagreement about
experimental evidence that quantum systems exist in a multitude
of states, with only a portion of those states being observable. The
approaches differ in what happens to the non-observed states
and in the process by which states become observed states.

The approach by Linde proposes that the universe follows
a deterministic evolution about which we can only observe partial
aspects of the multiple possible occurrences that are deterministically
created. The approach by Zurek proposes that the deterministic
evolution of the universe is constrained by a Darwinian extinction
of some of the possible evolution paths of the system. The approaches
by Linde and Zurek agree in what is observed, but they disagree
about what happens to the non-observed states. In the case of
quantum gravity effects during cosmic inflation these differences
may be relevant.

The random extinction of information in quantum Darwinism contrasts
with the preservation of information in Hilbert's formal axiomatic
systems (FAS) \cite{3}. The FAS is a deterministic system where the consistence
of the axioms is preserved, meaning that no proposition can be
both true and not true; and the logic in the FAS obtains true
propositions from true propositions. In Darwinian approaches
(e.g. quantum Darwinism) if survival is identified with truth,
then some true propositions lead to false propositions (which
become extinct) and so Darwinism is not consistent. However,
in Darwinism there are no true propositions that are not obtained
from true propositions (all entities have parent entities that
need to be true since they gave offspring); meaning Darwinism
is necessarily complete. G\"{o}del's incompleteness theorems showed
that a non-trivial FAS cannot be both complete and consistent \cite{3,11}.

An axiomatic system made to be complete and not consistent would
have validly inferred propositions being both true and not-true.
A form of dealing with this difficulty would be to validate propositions
not by the valid application of inference rules, but by using
a proof-checking algorithm that would eliminate propositions
that are inconsistent within themselves. Such a process of selecting
valid propositions is called here a Darwinian axiomatic system
(DAS), and is described more extensively in the Appendix. The FAS
and the DAS are the two extreme forms of dealing with G\"{o}del's
incompleteness theorems, respectively the consistent and the
complete forms. It is possible to conceive an hybrid axiomatic
system (HAS) between the FAS and the DAS.

This work mostly uses the works of Linde \cite{1} and Zurek \cite{2}; other
approaches considered are Chaitin's use of information in mathematics
\cite{3}, Wheeler's concept of `it for bit' \cite{4}, Rovelli's relational
approach to quantum gravity \cite{5}, Smolin's relation between quantum
mechanics and quantum gravity \cite{6}, and Guth's approach to cosmic
inflation \cite{7}.

Although other approaches to decoherence during cosmic inflation have been
developed \cite{Hu,Lombardo2002,Lombardo2004,Martineau,Burgess,Calzetta,Kiefer},
our approach differs
in that it is based on Zurek’s quantum Darwinism \cite{2} and does not rely
on having the short and long wavelength quantum fields
represent the environment and the system, respectively.

The remainder of the present article is structured as follows.
The next five sections discusses the underlying theory: Introduction
to Quantum Measurement; Introduction to Quantum Darwinism; Relation
between Quantum Darwinism and Quantum Diffusion; Diffusion in
Cosmic Inflation; Effects of a Quantum Darwinism Approach to
Cosmic Inflation. The calculations at the end of each subsection
are then presented in the section on Results, with their relations
highlighted. The section entitled Discussion and Summary describes
the possible implications of the results obtained and highlights
the principal results.

\section{Introduction to Quantum Measurement}

Quantum Darwinism \cite{2,10} is an approach to quantum measurement
that is strongly based on Wheeler's ``it-for-bit'' approach \cite{4}
and so it has parallels with both information theory and computation. The
classical technical definition of the amount of information was
provided by Shannon's information entropy and stated that if
the sending device has a probability $P_j$ of sending message \textit{j} from
a set of \textit{N} messages, then the information produced when one
message is chosen from the set is, in units of bits \cite{3},

\be
{\cal H}=-\log_{2} P_{j}\ .
\label{1}
\ee

For a brief description of quantum Darwinism it is helpful to
resort to a short description of the limitations of non-Darwinian
quantum mechanics, the limitations that quantum Darwinism addresses. In
quantum mechanics the universe is separable into 3 parts: I.
System  \textit{S,} II. Apparatus \textit{A,} III. Environment \textit{E}. The
evolution of quantum systems occurs according to Schr\"{o}dinger's
equation. Entanglement between system and apparatus can be modeled
by unitary Schr\"{o}dinger evolution. Von Neumann \cite{8} proposed
a non-unitary selection of the preferred
basis,

$$\left| \Psi _{SA} \right\rangle \left\langle \Psi _{SA} \right|
\rightarrow \sum\limits_{k}\left| a_{k} \right| ^{2} \left| s_{k}
\right\rangle \left\langle s_{k} \right|  \left| A_{k} \right\rangle
\left\langle A_{k} \right| =\rho _{SA}\ . $$

and also proposed the non-unitary ``collapse'' enabling
the occurrence of a unique outcome (e.g. for state 17):

$$\sum\limits_{k}\left| a_{k} \right| ^{2} \left| s_{k} \right\rangle
\left\langle s_{k} \right|  \left| A_{k} \right\rangle \left\langle A_{k}
\right| \rightarrow \left| a_{17} \right| ^{2} \left| s_{17} \right\rangle
\left\langle s_{17} \right| \left| A_{17} \right\rangle \left\langle
A_{17} \right|\ . $$

 Zurek \cite{2,10} proposed an approach to entanglement which is unitary
and as un-arbitrary as possible, using the environment. The use
of the environment implies abandoning the closed-system assumption \cite{10},
requiring the following alteration:

$$\left| \Psi _{SA} \right\rangle \left| e_{0} \right\rangle
=\left( \sum\limits_{k}a_{k} \left| s_{k} \right\rangle \left| A_{k}
\right\rangle  \right) \left| e_{0} \right\rangle \rightarrow
\sum\limits_{k}a_{k} \left| s_{k} \right\rangle \left| A_{k} \right\rangle
\left| e_{k} \right\rangle = \left| \Psi _{SAE} \right\rangle\ . $$

 The selection of the preferred basis is obtained using unitary
evolution by assuming $\left| \left\langle e_{k} \right.\left| e_{l} \right\rangle \right| ^{2}
=\delta _{kl} $ and tracing over the environment \cite{10},

$$\rho _{SA} =\mb{Tr}_{E} \left| \Psi _{SAE} \right\rangle \left\langle \Psi
_{SAE} \right| =\sum\limits_{k}\left| a_{k} \right| ^{2} \left| s_{k}
\right\rangle \left\langle s_{k} \right|  \left| A_{k} \right\rangle
\left\langle A_{k} \right|\ . $$

 The preferred basis is defined by the set of states the apparatus
can adopt that do not interact with the environment and therefore
only interact with the system. The apparatus adopts one of the
pointer states after it makes a measurement. For this set of
pointer states to exist it is necessary that the apparatus be
entangled with the environment. Entanglement is a non-classical
quantum behaviour where two parts of the universe that have interacted
at a certain point in time have to be described with reference
to each other even if they are now separated in space, as long
as they remain entangled. The above explanations of quantum measurement
do not clarify the meaning of tracing over the environment, and
the non-unitary ``collapse'' is not really explained. Quantum
Darwinism addresses both issues successfully \cite{2,12}.

\section{Introduction to Quantum Darwinism}

In quantum Darwinism, the following statements are considered
to be valid: (a) The universe consists of systems. (b) A pure
(meaning completely known) state of a system can be represented
by a normalized vector in Hilbert space H$_{\mathit{S}}$. (c) A composite
pure state of several systems is a vector in the tensor product
of the constituent Hilbert spaces. (d) States evolve in accordance
with the Schr\"{o}dinger equation $i\hbar | \dot{\psi } \rangle =H |\psi\rangle$
where $H$ is Hermitian. In quantum Darwinism no ``collapse'' postulate
is needed. An assumption by von Neumann \cite{8} and others is that
the observers acquire information about the quantum system from
the quantum system, but that is (almost) never the case. The
distinction between direct and indirect observation might seem
inconsequential as a simple extension of the von Neumann chain,
but the use of the point of view of the observer in quantum Darwinism
makes it possible to obtain the ``collapse" \cite{10,12}.

In quantum Darwinism there is ``no information without representation",
meaning that the information is always about a state that is
being represented. Preferred pointer states selected through
entanglement define what is being stored in the environment. The ``information amount"
in quantum systems is defined using the density matrix $\rho$ and is based on
ref. \cite{10}.

Environment-assisted invariance (envariance) is a quantum symmetry
exhibited by the states of entangled quantum systems. The joint
state can always be described by a Schmidt basis (if the environment
is made big enough).

\section{Relation between Quantum Darwinism and Quantum Diffusion}

In molecular Brownian motion there is a permanent oscillation
between position measurement and momentum measurement.
Brownian motion of quantum states describes decoherence; this
is also an accurate description of molecular Brownian motion. The
quantum Brownian motion model used here consists of an environment \textit{E}
made of a collection of harmonic oscillators of position \textit{q}$_{\mathit{n}}$,
mass \textit{m}$_{\mathit{n}}$, frequency \textit{w}$_{\mathit{n}}$, and coupling constant \textit{c}$_{\mathit{n}}$,
interacting with a system \textit{S} of mass \textit{M,} position \textit{x,} and
harmonic potential \textit{V(x)= \ensuremath{\frac12}MW}$^{\mathit{2}}$\textit{x}$^{\mathit{2}}$.
The total Lagrangian is \cite{13}

\be
L\left( x,q_{n} \right) =\underbrace{\frac{M}{2} \left[ \dot x^2-W^2x^2 \right]}_{L_{S}}
+\underbrace{\sum_n\fr{m_n}{2}\left[\dot q_n^2-w^2_n\left[q_n-\fr{c_nx}{m_nw_n^2}\right]^2\right]}_{ L_{SE}}\ .
\label{3}
\ee

The Lagrangian component \textit{L}$_{\mathit{SE}}$ takes into account the renormalization
of the potential energy. Let us denote $k$ as the
Boltzmann constant and $T$ as the temperature. If the thermal energy \textit{kT} is higher than
all other relevant energy scales, including the energy content
of the initial state and energy cutoff in the spectral density
of the environment \textit{C(v)}; then the master equation for the
density matrix \textit{\ensuremath{\rho}}$_{\mathit{S}}$ of an initially environment-independent
system \textit{S} depends on the renormalized Hamiltonian $H_{\mathit{ren}}$
and on

$$\gamma =\frac{2}{MW} \int\nolimits_{0}^{\infty
}dl\int\nolimits_{0}^{\infty }dv C\left( v\right)   \sin \left( Wl\right)
\sin \left( vl\right) $$

in the following way \cite{13}:

\be
\dot{\rho}_S= -\fr{i}{\hbar}[H_{ren},\rho_S]-\gamma[x-y]\left[\fr{\dl}{\dl x}-\fr{\dl}{\dl y}\right]
\rho_S-\fr{2M\gamma kT}{\hbar^2}[x-y]^2\rho_S\ .
\label{4}
\ee

 In this high \textit{T} case the master equation is independent of \textit{V(x)}.
The relaxation time is \ensuremath{\gamma}$^{\mathit{-1}}$ and the decoherence
time is \cite{10,9}:

\be
\tau _{D} =\gamma ^{-1} \left[ \frac{\frac{\hbar }{\sqrt{2MkT} } }{x-y}
\right] ^{2}\ .
\label{5}
\ee

The Wigner quasi-distribution representation \textit{Z} of the high
temperature density matrix master equation (Eq. (\ref{4})) is \cite{13}:

\be
\dot{Z}=-\fr{p}{M}\fr{\dl}{\dl x}[Z]+\fr{\dl V}{\dl x}Z+2\gamma\fr{\dl}{\dl p}[pZ]
+2\gamma MkT\fr{\dl^2}{\dl p^2}[Z]\ .
\label{6}
\ee

The minimum uncertainty Wigner quasi-distribution for a phase
space localized wave-packet is \cite{13}:
\be
Z\left( x_{0} ,x,p_{0,} p\right) =\frac{1}{\pi \hbar } \exp \left(
-\left[ \frac{x-x_{0} }{\sqrt{\frac{\hbar }{2MW} } } \right] ^{2} -\left[
\frac{p-p_{0} }{\sqrt{\frac{\hbar MW}{2} } } \right] ^{2} \right)\ .
\label{8}
\ee

If there are two wave packets separated by \textit{\ensuremath{\Delta}}\textit{x},
with average location \textit{x} and average momentum \textit{p}, then the
joint Wigner quasi-distribution is equal to averaging the two
localized Wigner distribution expressions plus a non-classical
interaction term equal to \cite{13}
\be
W_{\mbox{int} } \approx \frac{1}{\pi \hbar } \cos \left( \frac{\Delta
x}{\hbar } p\right) \exp \left( -\left[ \frac{x}{\sqrt{\frac{\hbar }{2MW}
} } \right] ^{2} -\left[ \frac{p}{\sqrt{\frac{\hbar MW}{2} } } \right]
^{2} \right)\ .
\label{9}
\ee

Joining the diffusion coefficient expression \cite{13,9}

$$D=\frac{kT}{\gamma M} $$

 with the decoherence-time definition of Eq. (\ref{5}) yields a relation
between decoherence-time and diffusion coefficient,

\be
\tau _{D} =\frac{D}{2} \left[ \frac{\hbar }{kT\left[ x-y\right] } \right]
^{2} \ .
\label{10a}
\ee

From Einstein's diffusion equation we know that $\langle (x(t)-x(0))^2\rangle=2Dt$ for a single molecule.
Consider now two molecules. Let $t_{\{x,y\}}$ be the time interval since the last collision of
two molecules which collided at the point $x_{0}=y_{0}$ and which are now at the positions $x$ and $y$,
respectively. Using the statistical independence of the two molecules,
$\langle x y \rangle=\langle x\rangle\langle y\rangle$ and noting that $\langle x\rangle=\langle y\rangle =x_{0}$,
the expression becomes
$\langle (x-y)^2\rangle=4Dt_{\{x,y\}}$. This is an expression for the {\it average} behavior of a pair of molecules.
A corresponding expression for the {\it particular} behavior of two molecules can be written as
$(x-y)^2=4D_{\{x,y\}}t_{\{x,y\}}$ where $D_{\{x,y\}}$ is a coefficient valid for that particular event. With the
reasonable assumption that this coefficient becomes $D_{\{x,y\}}\simeq D$ very fast, that is for any appreciable
distance $x-y$,\\
 Eq. (\ref{10a}) can now be rewritten as

\be
\tau _{D} =\frac{1}{8t_{\{x,y\}}} \left[ \frac{\hbar }{kT } \right]
^{2} \ .
\label{10b}
\ee

\section{Diffusion in Cosmic Inflation}

The purpose of this section is to describe how cosmic inflation
relates to Brownian motion. It is not intended to present a thorough
description of cosmic inflation. In the present description of cosmic
inflation there are multiple Big Bang occurrences, and in each
of these occurrences baby-universes are created \cite{1}. One of the baby-universes
is our own universe. In order to describe cosmic inflation
it is helpful to explain what is being inflated. The behavior
of spacetime is characterized by the relation between differences
in time and differences in spatial location, and can be represented
by the behavior of a single characteristic time varying scale parameter \textit{a} which
appears in the line element which is characteristic of spacetime.
The Hubble parameter is the fractional change of \textit{a} with time: $\mathbb{H}=\frac{1}{a} \frac{da}{dt} $.
 Inflation describes the early epoch period of rapid growth
of \textit{a}. During inflation $\mathbb{H}$ is approximately constant at a value roughly
of the order $\mathbb{H}\cong 10^{34} $s$^{-1} $
which makes \textit{a} approximately proportional to $\re^{\mathbb{H}t}$.
Inflation comes to an end when $\mathbb{H}$ begins to decrease rapidly.
The energy stored in the vacuum-like state is then transformed
into thermal energy, and the universe becomes extremely hot.
From that point onward, its evolution is described by the hot
universe theory.

To correctly describe Brownian behavior during cosmic inflation,
it is convenient to distinguish between two horizons: the particle
horizon and the event horizon. The particle horizon delimits
what an observer at a time \textit{t} can observe assuming the capacity
to detect even the weakest signals. The event horizon delimits
the part of the universe from which we can ever (up to some maximal
time \textit{t}$_{\mathit{max}}$) receive information about events taking place
now (at time \textit{t}). The particle and event horizons are in a
certain sense complementary. In an exponentially expanding universe,
the radius of the event horizon is equal to $c\mathbb{H}^{-1}$ where \textit{c}
is the speed of light in vacuum. In an exponentially expanding
universe, any two points that are more than a distance $c\mathbb{H}^{-1}$
apart will move away from each other faster than \textit{c}, meaning
that those two points will never observe each other. They might
belong to the same baby-universe if they come from the same Big
Bang, but the points will lie beyond each other's particle horizons.

As described in Ref. \cite{1},
cosmic inflation leads to the creation of multiple baby-universes
one of them our own. Some of those universes will have physical
behaviors very different from the behavior of our universe, but
we will now consider the behavior of quantum fluctuations in
the cosmic inflation model. The scalar inflaton field \textit{\ensuremath{\varphi}} (sometimes
identified with the Higgs field, although this is controversial) is represented
as \cite{1}

\be
\varphi \left( \textbf{x},t\right) =\left( 2\pi \right) ^{-\frac{3}{2} } \int
d^{3} p \left[ a_{p}^{+} \psi _{p} \left( t\right) e^{i\textbf{px}} +a_{p}^{-} \psi
_{p}^{*} \left( t\right) e^{-i\textbf{px}} \right] \ .
\label{11}
\ee

 The $\left( 2\pi \right) ^{-\frac{3}{2} } $ term is simply a normalization factor,
$\int d^{3} p $ is the integration over all possible values of the momentum, $a_{p}^{+} $
creates a field with momentum \textbf{p} parameter with a probability
modulated by $\psi _{p} \left( t\right) $ and propagating in spacetime as the wave $e^{i\textbf{px}} $, and
$a_{p}^{-} $ destroys that same field.

The first cosmic inflation models considered that \textit{\ensuremath{\varphi}}
was a classical field (meaning non-quantum). The way a quantum
system becomes classical is through the process of decoherence. As
described in the previous section, the process of decoherence
has strong similarities to Brownian motion.  Ref. \cite{1} describes the
similarity of the behavior of \textit{\ensuremath{\varphi}} during cosmic inflation
and Brownian motion.

As it is typical in Brownian motion, the diffusion of the field \textit{\ensuremath{\varphi}}
can be described by the probability distribution $P\left( \varphi ,t\right) $
 of finding the field \textit{\ensuremath{\varphi}} at that point in instant \textit{t}.  In
Eq. 7.3.17 of Ref. \cite{1} it is found that

\be
\frac{\partial P\left( \varphi ,t\right) }{\partial t} =D\frac{\partial
^{2} P\left( \varphi ,t\right) }{\partial \varphi ^{2} } \ .
\label{12}
\ee

 Using Eq. (\ref{12}), Ref. \cite{1} shows that

$$\langle \varphi ^{2} \rangle=2Dt$$

 as is expected in diffusion processes (Eq. 7.3.12 in Ref. \cite{1}) and
that

\be
D=\frac{\mathbb{H}^{3} }{8\pi ^{2} c^{2} }\ .
\label{13}
\ee

\section{Effects of a Quantum Darwinism Approach to Cosmic Inflation}

The way a quantum system becomes classical is through the process
of decoherence, which according to quantum Darwinism is described
by quantum Brownian motion in the high temperature limit. So it
is possible that the Brownian process in cosmic inflation described
in Ref. \cite{1} entails the extinction of the non-decohered universe
states.

G\"{o}del's incompleteness theorems propose to describe the difficulties
of creating a mathematical formalism from nothing using Hibert's
FAS \cite{3,11}, which is a deterministic approach. Quantum Darwinism proposes to
address the creation of classical reality from a quantum reality, using
a Darwinian approach. The deterministic and the Darwinian approach to
creation can be considered as the two extreme approaches of dealing
with G\"{o}del's incompleteness theorems (see Appendix). The Big Bang
proposes to describe the creation of an observable universe from
nothing, and so it will be very Darwinian. A Darwinian evolution is a Brownian evolution
where extinction might occur; and so this study of the relation
between decoherence (extinction of some quantum states) and diffusion
(Brownian motion) of baby-universes is a study of Darwinian processes
occurring during cosmic inflation.

Solving the diffusion equation (\ref{12}) during cosmic inflation,
one obtains the probability for creation of a universe with a
certain vacuum energy. Summing over all topologically disconnected
configurations of just-created universes enables obtaining the probability
for creating universes with a certain cosmological constant value \cite{1},
causing Linde to write that although ``\textit{it is often supposed
that the basic goal of theoretical physics is to find exactly
what Lagrangian or Hamiltonian correctly describes our entir}\textit{e
world. \ldots one could well
ask \ldots if the
concept of an observer may play an important role not just in
discussions of the various characteristics of our universe, but
in the very laws by which it is governed}.'' The answer proposed
here to Linde's question is that if the quantum Darwinism approach
is applied to cosmic inflation, then the laws of physics are
themselves the result of a Darwinian evolution of quantum systems.

\section{Results}

We use Eq. (\ref{10b}) to generalize the results obtained for molecules
in quantum Brownian motion to baby-universes undergoing Brownian
motion during cosmic inflation. The decoherence time $\tau_D$ is then a time duration referring to
two baby-universes, with $t$ being the time since they last interacted
(typically the last time they were at the same place would be at the beginning of the Big Bang).
The decoherence time is obtained as (for an expression in terms of $\varphi$ see footnote \footnote{In a ``metaphorical" approach to quantum field theory
the position $x$ is replaced by the field $\phi$ and the
momentum $p$ by the field $\psi$ \cite{1}. Then the time during which two baby-universes -
located at \textit{x} and \textit{y} and described
by the corresponding fields $\varphi(x)$ and $\varphi(y)$,
respectively - remain entangled after initial interaction would be
$\tau _{D} =\mb{H}^{-1} \left[ \frac{\hbar }{4\pi ck} \frac{\mathbb{H}^{2} }{T\left[
\varphi(x) -\varphi(y) \right] } \right]
^{2}\ .$}):

\be
\tau _{D} =\frac{1}{8t} \left[ \frac{\hbar }{kT} \right]
^{2} \ .
\label{14}
\ee

The differing approaches by Linde and Zurek, which can be linked
to the differences between the axiomatic systems FAS and DAS,
imply different outcomes for the non-observed states. The FAS/Linde approach considers
that the outcomes incompatible with the observed outcome exist
in different multi-verses, while the DAS/Zurek approach considers
the outcomes incompatible with the observed outcome to have become
non-existent. In Zurek's approach information-transmission is
what enables existence \cite{13}, while in quantum gravity existence
(expressed as the number of quantum particles) is observer dependent and
thus can only be understood as a relational concept \cite{1,5}.

The representation of cosmic inflation using a diffusion process
in a de Sitter space allows to consider thermal equilibrium with \cite{14,1}

$$T=\frac{\hbar \mathbb{H}}{k} $$

 so that Eq. (\ref{14}) becomes

\be
\tau _{D} =\frac{1}{8t} \left[ \frac{1}{\mathbb{H}} \right]
^{2} \ .
\label{15}
\ee

This result implies that during the duration of cosmic inflation, the decoherence
time is much smaller than the cosmic inflation duration (for an estimate see footnote \footnote{In cosmic inflation, there is a growth by typically at
least a factor
of e$^{60}$, starting at $t=10^{-35}$s and ending at $t=10^{-32}$s, so that $\Delta t$ is about $10^{-32}$s. Therefore,
since $\mathbb{H} \Delta t = 60$, the Hubble parameter $\mathbb{H}$ is about $10^{34}$s$^{-1}$,
and the expression $1/(8 H^2 t)$ is about $10^{-36}$s.
Compared with $10^{-35}...10^{-32}$s this is much smaller.}).  Meaning that,
baby-universes will be in a quantum coherent state for only a small fraction of the
duration of cosmic inflation.  This result agrees with Martineau's result \cite{Martineau} that
decoherence is extremely effective during inflation, but reaches that result
in a much more simple way.  The approach to ``decoherence during Brownian motion"
used by Zurek considers that the effect of zero-point vacuum fluctuations is neglected.
Kiefer et al. \cite{Kiefer} propose that the inclusion of zero-point vacuum fluctuations makes
decoherence still effective but no longer complete, meaning that a significant part of
primordial correlations remains up to the present moment.

\section{Discussion and Summary}

Obtaining values for the decoherence time requires knowledge
of the value of the Hubble parameter before and during inflation. Values
of the Hubble parameter have a large range, and the measurement of
its value is a topic of current research \cite{1}. The existence
of baby-universes is also a not yet established observational
fact \cite{1}. Thus, obtaining experimental proof of Eq. (\ref{14}) and Eq. (\ref{15}) is not
yet possible. But if baby-universes exist, and if more information
is obtained about the time-dynamics of the Hubble parameter,
the relation between Hubble parameter and decoherence-time expressed
in Eq. (\ref{14}) and Eq. (\ref{15}) would be likely to become useful.

A characteristic of biological Darwinism is the existence of
a first cell.  The approach to cosmic inflation described in Ref. \cite{15} indicates that the
inflating region of spacetime must have a past boundary; this
truly initial Bang would have occurred a lot earlier than our
own Big Bang. In this work a relation between Quantum Darwinism
and HAS is presented in the Appendix, with the HAS becoming more
and more Darwinian as the forces considered become closer to
what they were at the truly initial Bang (the initial forces,
because of their extremely high energy, are likely to be also
the most fundamental forces). The Quantum Darwinism treatment
of the truly initial Big Bang would therefore correspond to a
process that is as Darwinian as it gets, even more Darwinian
than the evolution of species.  This Darwinian process would become
more and more deterministic as the interactions between aspects of the
universe \textit{de facto} measure those aspects of the universe.
The measurement described in Eq. (\ref{14}) and Eq. (\ref{15}) obtains what
the physical constants (and laws) will be for a certain baby-universe
by a Darwinian extinction of the other possible values. That
the measurement occurring during cosmic inflation is the selector
of the physical constants is already proposed in section 10 of
Ref. \cite{1}, but the approach proposed here is different in that
it proposes the Darwinian extinction of the non-obtained quantum
alternatives that are not moving away at a speed faster than \textit{c}.

To summarize, an expression was obtained for the time after which different
previously entangled baby-universes would decohere.

\begin{acknowledgments}

NFL thanks Nachum Dershowitz, Jean-Claude
Zambrini, Juan Sanchez, and Juan Pablo Paz for answering his questions.
This work was partially supported by Fundação para a Ciência e a Tecnologia: Programa Ciência 2007 and
CERN POCI/FP/81926/2007.

\end{acknowledgments}

\appendix*

\section{Formal and Darwinian Axiomatic Systems}

A FAS is constituted by alphabet, grammar, axioms, rules of inference,
and a proof-checking algorithm. In the FAS approach to mathematics,
one starts with axioms considered as self-evident and built using the
alphabet and the grammar; then the rules of inference are
applied to the axioms and all the theorems (logical inferences
of the axioms) are obtained. A proof-checking algorithm checks
if a proof follows all the rules by doing reverse inference starting
from the proof's result and checking if what is obtained are
the axioms. G\"{o}del's incompleteness theorems showed that a non-trivial
FAS cannot be both complete and consistent \cite{3,11}.

Axioms in FAS are typically made to be consistent so that the
FAS is consistent, but an FAS cannot be both consistent and complete.
A form of dealing with this difficulty is to validate propositions
not by the valid application of inference rules, but by using
a proof-checking algorithm that would eliminate propositions
that are inconsistent within themselves. Such a process of selecting
valid propositions is called here a Darwinian axiomatic system
(DAS). The FAS and the DAS are the two extreme forms of dealing
with G\"{o}del's incompleteness theorems, respectively the consistent
and the complete forms. It is possible to conceive an hybrid
axiomatic system (HAS) which lies in-between the FAS and the DAS. In the next
paragraphs it will be proposed that Quantum Darwinism is similar
to an HAS.

To Chaitin's information-based G\"{o}del incompleteness conclusion \cite{3}
that real numbers are non-computable with probability 1, quantum
Darwinism answers through a discrete universe. In mathematical
randomness \cite{3} the value of a random variable is only known by
running a computer, and in quantum Darwinism the value of a random
quantum variable only occurs if the interaction in an experiment
is strong enough \cite{12}. The quantum randomness \cite{10,12} concept
is identical to the mathematical randomness \cite{2} concept if the
quantum systems' existence is enabled through their transmission
of information, which occurs in quantum Darwinism. The existence's
dependence on information is the part of the Existentialist philosophical
structure added by quantum Darwinism \cite{10}.

Quantum Darwinism's Existentialism does not allow for an actual
computation as that expressed in Turing's halting problem both
because of its lack of concrete existence and its lack of absolutely
closed systems. It is proposed here that envariance \cite{12} enabled
by quantum entanglement is an expression of Turing's halting
problem in quantum Darwinism. The entanglement between system
and its environment means that a program running on the system
can be counter- run by a program in the environment.



\begin{thebibliography}{15}

\bibitem{1}
A. Linde, hep-th/0503203v1.

\bibitem{2}
W.H. Zurek, quant-ph/0308163v1.

\bibitem{3}
G.J. Chaitin, \textit{Meta Math!}, (Random House, New York, 2005).

\bibitem{4}
J.A. Wheeler, in \textit{Proceedings of the 1$^{st}$ International
Sakharov Conference on Physics}, editors L.V. Keldysh and V.Y. Feinberg, Vol. 2 (1992).

\bibitem{5}
C. Rovelli, \textit{Quantum Gravity}, (Cambridge University
Press, Cambridge, 2004).

\bibitem{6}
L. Smolin, \textit{Three Roads to Quantum Gravity}, (Perseus Books Group, New York, 2002).

\bibitem{7}
A.H. Guth, Phys. Rev. \textbf{D23}, 347 (1981).

\bibitem{Hu}
B.L. Hu, J.P. Paz, Yuhong Zhang, gr-qc/9512049.

\bibitem{Lombardo2002}
F.C. Lombardo, F.D. Mazzitelli, R.J. Rivers, Nucl.Phys. B672 (2003) 462 [hep-ph/0204190].

\bibitem{Lombardo2004}
F.C. Lombardo, Braz.J.Phys. 35 (2005) 391 [gr-qc/0412069].

\bibitem{Martineau}
P. Martineau, astro-ph/0601134.

\bibitem{Burgess}
C.P. Burgess, R. Holman, D. Hoover, astro-ph/0601646.

\bibitem{Calzetta}
E. Calzetta, E. Verdaguer, J.Phys. A39 (2006) 9503 [quant-ph/0603047].

\bibitem{Kiefer}
C. Kiefer, I. Lohmar, D. Polarski, A. A. Starobinsky, Class. Quantum Grav. 24 (2007) 1699 [astro-ph/0610700].

\bibitem{8}
J. von Neumann, \textit{Mathematical Foundations of Quantum Mechanics},
(Princeton University Press, Princeton, 1955).

\bibitem{9}
D.A. McQuarrie, \textit{Statistical Mechanics}, (Harper
\& Row, New York, 1975).

\bibitem{10}
W.H. Zurek,  \textit{Decoherence and the transition from quantum
to classical - revisited}, Los Alamos Science {\textbf 27}, 86 (2002).

\bibitem{11}
K. G\"{o}del, Monatshefte f\"{u}r Mathematik und Physik {\textbf 38}, 173 (1931).

\bibitem{12}
W.H. Zurek, quant-ph/0405161v2.

\bibitem{13}
W.H. Zurek,  Rev. Mod. Phys. {\textbf 75}, 715 (2003).

\bibitem{14}
G.W. Gibbons  and S.W. Hawking, Phys. Rev. {\textbf D15}, 2752 (1977).

\bibitem{15}
A. Borde, A.H. Guth, and A. Vilenkin, Phys. Rev. Lett. \textbf{90}, 151301 (2003).

\end{thebibliography}
\end{document}